\documentclass[conference,hidelinks,a4paper]{IEEEtran}
\IEEEoverridecommandlockouts
\usepackage{cite}
 \usepackage{color}

\usepackage{float}
\usepackage{amsmath,amssymb,amsfonts}
\usepackage[english]{babel}
\usepackage{amsthm}
\theoremstyle{definition}

\usepackage{graphicx}
\usepackage{caption}
\usepackage{subcaption}
\usepackage{multirow}
\usepackage[left=1.35cm,right=1.35cm,top=1.9cm,bottom=4.1cm]{geometry}
\usepackage{textcomp}
\usepackage{xcolor}
\usepackage{bbm}
\usepackage{algorithm}
\usepackage{algpseudocode}
\def\BibTeX{{\rm B\kern-.05em{\sc i\kern-.025em b}\kern-.08em
    T\kern-.1667em\lower.7ex\hbox{E}\kern-.125emX}}
\usepackage{tikz}
\begin{document}
\def\blue{\textcolor{blue}}
\title{GNN-Based Joint Channel and Power Allocation in Heterogeneous Wireless Networks\\

}
\newcommand\blfootnote[1]{%
  \begingroup
  \renewcommand\thefootnote{}\footnote{#1}%
  \addtocounter{footnote}{-1}%
  \endgroup
}
\author{\IEEEauthorblockN{Lili Chen, Jingge Zhu and Jamie Evans}
\IEEEauthorblockA{Department of Electrical and Electronic Engineering, University of Melbourne, Australia \\
Email: lilic@student.unimelb.edu.au, jingge.zhu@unimelb.edu.au, jse@unimelb.edu.au\\
}

}

\maketitle

\begin{abstract}
 The optimal allocation of channels and power resources plays a crucial role in ensuring minimal interference, maximal data rates, and efficient energy utilisation. As a successful approach for tackling resource management problems in wireless networks, Graph Neural Networks (GNNs) have attracted a lot of attention. This article proposes a GNN-based algorithm to address the joint resource allocation problem in heterogeneous wireless networks. Concretely, we model the heterogeneous wireless network as a heterogeneous graph and then propose a graph neural network structure intending to allocate the available channels and transmit power to maximise the network throughput. Our proposed joint channel and power allocation graph neural network (JCPGNN) comprises a shared message computation layer and two task-specific layers, with a dedicated focus on channel and power allocation tasks, respectively. Comprehensive experiments demonstrate that the proposed algorithm achieves satisfactory performance but with higher computational efficiency compared to traditional optimisation algorithms.
  
\end{abstract}


\section{Introduction}\label{sec:introduction}
In the ever-changing landscape of wireless communication systems, the demand for high-quality, reliable, and efficient data transmission has escalated significantly. This increasing demand is driven by countless applications, including mobile communications, the Internet of Things, smart cities, and autonomous vehicles. Properly allocating available channels and power resources is fundamental to addressing these challenges and unlocking the full potential of wireless systems. Many wireless communication standards and technologies, such as 5G and beyond, Wi-Fi, and emerging wireless ad hoc networks, are designed to accommodate a growing number of users and devices. In this context, the optimal allocation of channels and power resources plays a crucial role in ensuring minimal interference, maximal data rates, and efficient energy utilisation. Moreover, the efficient allocation of these resources holds the promise of enabling groundbreaking applications, such as low-latency communications for mission-critical tasks, ultra-reliable communication, and connectivity in highly dense environments.
\blfootnote{The work was supported by the Melbourne Research Scholarship of the University of Melbourne and in part by the Australian Research Council under projects DE210101497 and DP220103281.}

Several studies have already highlighted the importance of channel allocation and power allocation. The authors in \cite{feng2013device} present a structure for joint channel and power allocation (JCPA) in Device-to-Device (D2D) communication underlying cellular networks to maximise the network sum rate while under several quality of service constraints. They proposed a three-step scheme in which Quality of Service (QoS) aware admission and power control are performed first, followed by the utilisation of a maximum weight bipartite matching algorithm to determine the optimal channel allocation for maximising the overall network throughput. A similar maximum weight bipartite matching algorithm is also demonstrated in \cite{meshgi2015joint} for multicast D2D communication.

In the problem formulations outlined in \cite{feng2013device} and \cite{meshgi2015joint}, it is stipulated that each D2D pair is restricted to the utilisation of a singular subchannel, and likewise, each channel can be accessed by no more than one D2D pair. Nevertheless, this approach, characterised by a subchannel reuse mode, is deemed suboptimal in terms of spectrum efficiency. Therefore, the investigation conducted by the authors in \cite{zhao2015resource} is centred on addressing this issue by allowing two D2D pairs to share a common subchannel while ensuring QoS. To take a step further, the framework for multiple D2D users to share the same subchannel is proposed in \cite{abdallah2018power} to maximise network throughput. However, the traditional optimisation methods are unsuitable for numerous practical application scenarios due to their high complexity.

To reduce the computational time while achieving good performance, machine learning is widely applied to wireless communication problems. Graph Neural Networks (GNN), with the ability to exploit the topology information of wireless networks, have been utilised in many resource allocation problems\cite{gao2022decentralized,10283511,he2020resource}. Channel management with GNNs in wireless networks is considered in \cite{gao2022decentralized} and \cite{nakashima2020deep}, aiming to improve user experiences and minimise mutual interference among access points, respectively. Similarly, \cite{he2020resource} used GNN to extract the features of Vehicle to Vehicle pairs, subsequently employing these feature vectors in the reinforcement learning framework to maximise the network throughput. To optimise the channel and power allocation together, the authors in \cite{chen2021gnn} proposed GNN to learn the channel allocation when transmit power is fixed. Then, optimal power allocation is derived based on fixed channel allocation. Nonetheless, their methodology imposes a restriction wherein a channel can be accessed by at most one D2D pair. Furthermore, their proposed solution exhibits limitations when applied to scenarios with larger-scale problems, as the algorithm relies on labelled training samples with high computational complexity.

To address these limitations, this paper delves into the design of end-to-end GNN frameworks for the joint optimisation of channel and power allocation in heterogeneous wireless networks. Our approach leverages GNNs to exploit topological information of the heterogeneous wireless networks and extract relevant channel and interference features. Different from the approach presented in \cite{chen2021gnn}, our GNN design allows for the concurrent access of a channel by multiple D2D pairs, facilitating a reuse mode that aligns well with numerous practical applications. Moreover, our algorithm exhibits benefits by eliminating the dependence on labelled training data and demonstrates applicability in large-scale network scenarios.

\section{Resource Allocation problems}
In this subsection, we present a system model and formulate the joint channel and power allocation as an optimisation problem. 
\subsection{System Model}
\begin{figure}[htbp]
\centerline{\includegraphics[width=9cm]{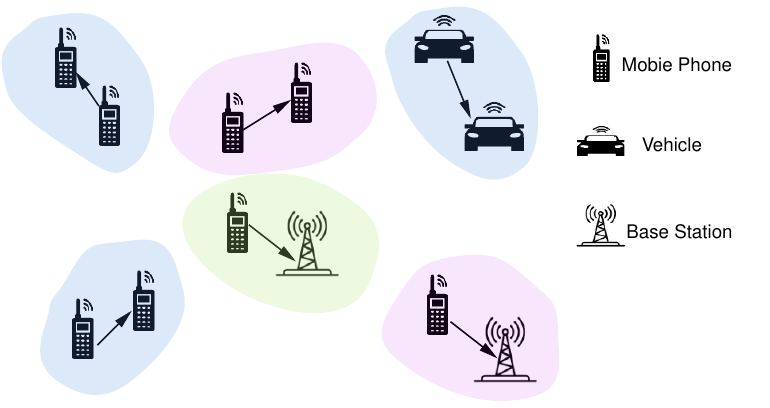}}
\caption{Heterogeneous wireless network with different channels.}
\label{fig:systemodel}
\end{figure}

Consider a heterogeneous wireless network as shown in Figure~\ref{fig:systemodel} with $D$ transceiver pairs denoted by $\mathcal{D} = \{1,2,...,D\}$, where each transmitter or receiver can be a vehicle, base station and mobile phone. The transceiver pairs exhibit varying performance requirements and possess the flexibility to select from accessible radio resources, i.e., channels and power, to meet these demands. In Figure~\ref{fig:systemodel}, we employ different colours to indicate different channels.  Mutual interference arises when two transceiver pairs share the same channel.
There are $M$ orthogonal channels, each with identical bandwidth, to be used in the system. We represent the index set for these channels as $\mathcal{M} = \{1, 2, ..., M\}$. In this system model, we follow a similar set-up in \cite{nakashima2020deep} that we refrain from assigning specific values to the bandwidths of the channels, but presume the absence of overlap between them. The received signal at the $i$-th receiver in $m$-th channel is given by 

\begin{equation}
    y_{i}^{m} = h_{i,i}^{m} s_{i} c_{i}^{m}+\sum_{j \neq i}h_{i,j}^{m} s_{j} c_{j}^{m}+n_{i},
\end{equation}
where $h_{i,i}^{m} \in \mathbb{C}$ is the channel state information (CSI) between $i$-th transceiver pair and $h_{i,j}^{m} \in \mathbb{C}$ is the interference between transmitter $j$ and receiver $i$ at $m$-th channel. The transmitted signal for transmitter $i$ is represented by $s_{i} \in \mathbb{C}$. Let $c_{i}^{m}$ denote the indicator variable of the channel allocation, where $c_{i}^{m}=1$ if $m$-th channel is utilised by $i$-th transceiver pair, and $c_{i}^{m}=0$ otherwise. The additive Gaussian noise at receiver $i$ is modelled as $n_{i} \sim \mathcal{C} \mathcal{N}\left(0, \sigma_{i} ^2 \right)$. The signal-to-interference-plus-noise ratio (SINR) of the receiver $i$ at $m$-th channel is expressed as follows,
\begin{equation}
    \text{SINR}_{i}^{m} = \frac{\left|h_{i,i}^{m}\right|^{2} p_{i} c_{i}^{m}}{\sum_{j \neq i}\left|h_{i,j}^{m}\right|^{2} p_{j} c_{j}^{m}+\sigma_{i}^{2}}
\end{equation}
where $p_{i}$ is the power for $i$-th transmitter. Denote $\mathbf{C} = [\mathbf{c}_1,...,\mathbf{c}_D]^{T} \in \mathbb{R}^{D \times M}$ where $\mathbf{c}_i = [c_{i}^1,...,c_{i}^{M}]$ is a one-hot vector and $\mathbf{P} = [p_1,...,p_D]^{T} \in \mathbb{R}^{D \times 1}$. 
Assume the channel bandwidth is equally distributed among all the channels, therefore, the data rate for $i$ receiver at $m$-th channel is given by,
\begin{equation}
R_{i}^m(\mathbf{C,P})=\log_2 \left(1+\frac{\left|h_{i,i}^{m}\right|^{2} p_{i} c_{i}^{m}}{\sum_{j\neq i}\left|h_{i,j}^{m}\right|^{2} p_{j} c_{j}^{m}+\sigma_{i}^{2}}\right)
\end{equation}

By adaptively allocating channels $\mathbf{C}$ and power $\mathbf{P}$ among transceiver pairs, our objective is to maximise the network throughput within the system while adhering to the specified maximum power constraint. The problem is formulated as,
 \begin{equation}
    \begin{array}{cl}
    \underset{\mathbf{C,P}}{\operatorname{maximise}} & \sum_{m=1}^{M}  \sum_{i=1}^{D}  w_{i} R_{i}^m(\mathbf{C,P}), \\
    \text { subject to } & 0 \leq p_{i} \leq p_{\max }, \quad  \forall i \in \mathcal{D},\\
    & c_{i}^{m} \in\{0, 1\}, \quad \forall i \in \mathcal{D}, m \in \mathcal{M}\\
    &\sum_{m=1}^{M} c_{i}^{m}=1, \quad \forall i \in \mathcal{D},
    \end{array}
    \label{eq:jointchannel}
\end{equation}   
where $w_{i}$ is the weight for $i$-th transceiver pair and $p_{\max }$ denotes the maximum power for $i$-th transmitter. 

Due to the interaction between optimisation variables, JCPA are normally considered to be non-convex problems and difficult to solve. To overcome this challenge, in the subsequent section, we introduce an efficient GNN-based algorithm to solve the joint resource allocation problem outlined in (\ref{eq:jointchannel}).

\section{graph neural networks for JCPA problems}
In this section,  we model the heterogeneous wireless network as a heterogeneous graph and then propose a GNN-based algorithm intending to allocate the available channels and transmit power to maximise the network throughput.

\subsection{Graph Modelling in Wireless Networks}

Graphs provide a method to model abstract concepts of wireless networks, such as the relationships between transmitter and receiver. A graph $\boldsymbol{G} = (\mathcal{V},\mathcal{E})$ can be characterised by the set of vertices $\mathcal{V}$ and edges $\mathcal{E}$ of a graph. For any given vertex $i,j\in\mathcal{V}$, $e_{i, j}$  defines the edge between them. The adjacency matrix of a graph is denoted as $\mathbf{A}$ and is an $n \times n$ matrix with entries in $\{0, 1\}$, where $\mathbf{A}_{i,j}=1$ if and only if $e_{i, j} \in \mathcal{E}$ for all $i,j\in \mathcal{V}$.
\begin{figure}[htbp]
\centerline{\includegraphics[width=9cm]{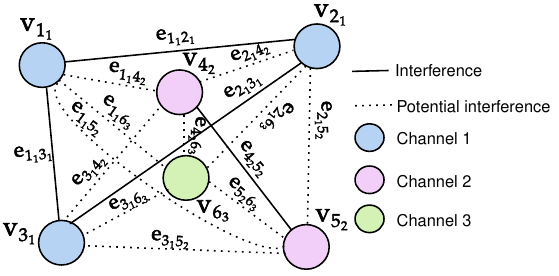}}
\caption{Heterogeneous graph representation of the system model.}
\label{fig:graphrepre}
\end{figure}

In our problem, heterogeneous wireless networks can be effectively modelled as heterogeneous graphs, where different types of transceiver pairs are represented as distinct nodes. Specifically, vehicular pairs, mobile user pairs, and cellular user pairs in these networks are treated as vertices, while edges are introduced to represent interference between them. To elaborate, given the interference present if transceiver pairs share the same channel, we introduce an interference edge between them. Furthermore, considering the dynamic nature of channel allocation in our model, additional edges, termed potential interference edges, are introduced. These edges signify the potential for interference between transceiver pairs if they were to share the same channel in subsequent iterations. An illustration of the graph representation is shown in Figure~\ref{fig:graphrepre}.

We define vertex as $i_m$ if $i$-th transceiver is allocated to $m$-th channel. Since interference only occurs if the transceiver pairs are in the same channel, for any vertex $i,j\in\mathcal{V}$, we define the interference edge between them as $e_{i_m, j_m}$. Similarly, we define $e_{i_m, j_n}$ as the potential interference edge if $m\neq n$ for any $i,j\in \mathcal{V}$. Denote the neighbouring set and the potential neighbouring set of vertex $i_m$ as $\mathcal{N}(i_m) = \{j_m|j\in\mathcal{V},j\neq i\}$ and $\mathcal{N}^p(i_m) = \{j_n|j\in\mathcal{V}, j\neq i,m\neq n\}$ respectively. The node feature incorporates the properties of the transceiver pairs, e.g., direct channel information and the transceiver pair's weight. We denote the node feature vector of vertex $i_m$ by $\mathbf{v}_{i_m} = \left[\left|h_{i, i}^m\right|, w_{i}\right]$,
where $h_{i, i}^m$ the CSI between $i$-th transceiver pair in $m$-th channel and $w_{i}$ is the corresponding weight. Since the edge feature represents the properties of the interference channel, we denote the interference edge feature vector  $\mathbf{e}_{i_m,j_m}=[|h_{i, j}^m|,|h_{j, i}^m|]$ and potential interference edge feature vector $\mathbf{e}_{i_m,j_n}=[|h_{i, j}^m|,|h_{i, j}^n|]$. 
With all the definitions in place, in the next section, we introduce a novel graph neural network structure designed to jointly address the power and channel allocation problem.
\subsection{JCPA on Heterogeneous Graph Neural Networks}
With wireless network modelled as a heterogeneous graph $\boldsymbol{G}$, the goal is to find a function $f_{\theta}(\cdot)$ mapping the $\boldsymbol{G}$ to the optimal channel allocation $\hat{\mathbf{C}}$ and power allocation $\hat{\mathbf{P}}$, where $\theta$ are learnable parameters. Message-passing graph neural networks (MPGNN) are introduced in \cite{shen2020graph} to tackle radio resource management problems. 
The update rule of MPGNN in the $s$-th layer at vertex $i$ is given by
\begin{equation}
\boldsymbol{h}_{i}^{(s)}=\alpha^{(s)}\left(\boldsymbol{h}_{i}^{(s-1)}, \phi^{(s)}\left\{\left[\boldsymbol{h}_{j}^{(s-1)}, \boldsymbol{e}_{j, i}\right]: j \in \mathcal{N}(i)\right\}\right),
    \label{eq:shenGNN}
\end{equation}
where $\boldsymbol{h}_{i}$ is the embedding vector for vertex $i$, $\boldsymbol{e}_{j, i}$ is the edge feature, $\mathcal{N}(i)$ is the neighbour of vertex $i$, $\phi^{(s)}(\cdot)$ is aggregation function that aggregate the information from vertex $i$'s neighbour and $\alpha^{(s)}(\cdot)$ is the update function that updates the embedding vector by combining aggregated information from its neighbours and its own information. Typically, Multilayer Perceptron (MLP) is used in aggregation and update functions due to universal approximation capacity \cite{hornik1989multilayer}.

\begin{figure*}[htbp]
\centerline{\includegraphics[width=15cm]{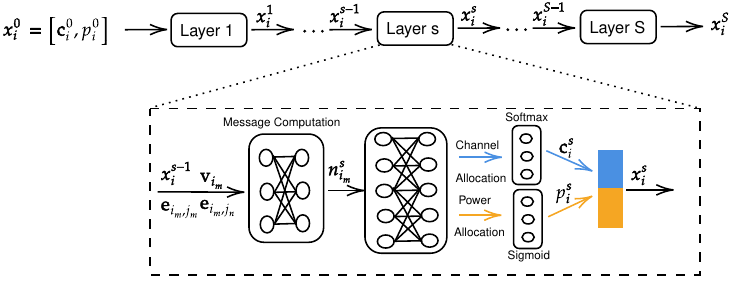}}
\caption{The structure of JCPGNN.}
\label{fig:GNNstructure}
\end{figure*}

Despite the success of MPGNN, it is difficult to solve the JCPA problems due to the shared aggregation and update function. This means that the embedding vector is the same for both problems and results in unsatisfied performance. In order to solve the problems, we propose a joint channel and power allocation graph neural network (JCPGNN) structure that generates two separate embedding vectors for each task. As shown in Figure~\ref{fig:GNNstructure}, JCPGNN contains two parts: 1) the shared message computation layer and 2) the task-specific layer. The message computation layer extracts the wireless network information and then passes it to the task-specific layer. Then, the task-specific layer converts this information to the estimated resource allocations. In our problem, each task-specific layer outputs the channel matrix $\mathbf{C}$ and power vector $\mathbf{P}$, respectively. The structure of JCPGNN is shown in Figure~\ref{fig:GNNstructure}.
\subsubsection{Message Computation Layer}
Message computation is a basic unit for graph neural networks. Since the channel and power allocation are interactive with each other, we use the shared message computation layer to extract the information from the transceiver pair. The update rule for message computation in the $s$-th layer at vertex $i$ is given by
\begin{equation}
\begin{aligned}
\boldsymbol{m}_{i_m,j_m}^{(s)}&=\phi_1^{(s)}\left\{\left[\boldsymbol{x}_{i}^{(s-1)},\mathbf{v}_{i_m}, \mathbf{e}_{i_m, j_m}\right]: j_m \in \mathcal{N}(i_m)\right\},\\
\boldsymbol{m}_{i_m,j_n}^{(s)}&=\phi_1^{(s)}\left\{\left[\boldsymbol{x}_{i}^{(s-1)},\mathbf{v}_{i_m}, \mathbf{e}_{i_m, j_n}\right]: j_n \in \mathcal{N}^p(i_m)\right\},\\
\boldsymbol{n}_{i_m}^{(s)} &=\phi_2^{(s)}\left\{\left[\boldsymbol{m}_{i_m,j_m}^{(s)},\boldsymbol{m}_{i_m,j_n}^{(s)}\right]\right\},
\label{eq:aggreated}
\end{aligned}
\end{equation}

where $\boldsymbol{x}_{i} = [\mathbf{c}_{i},p_{i}]$ is the optimisation variable. 
 
In the first step, a nonlinear transformation $\phi_1^{s}(\cdot)$ (e.g., MLPs) is applied to the information from the neighbour and itself $\boldsymbol{x}_{i}$ to extract the messages $\boldsymbol{m}_{i_m,j_m}$ and $\boldsymbol{m}_{i_m,j_n}$. In the second step, an aggregation function $\phi_2^{s}(\cdot)$ (e.g., SUM operation) is utilised to collect all the messages and output aggregated messages $\boldsymbol{n}_{i_m}$.

\subsubsection{Task-specific Layer}
Since channel $\textbf{C}$ and power allocation $\textbf{P}$ are different variables, it is better to use different update functions to generate them. Therefore, inspired by \cite{zhang2022learning}, we introduce a task-specific layer that includes two update functions in our proposed framework. The update rule for task-specific in the $s$-th layer at vertex $i$ is given by
\begin{align}
\mathbf{c}_{i}^{(s)}&=\alpha_1^{(s)}\left(\boldsymbol{x}_{i}^{(s-1)},\boldsymbol{n}_{i_m}^{(s)} \right),\label{eq:channel} \\
p_{i}^{(s)}& =\alpha_2^{(s)}\left(\boldsymbol{x}_{i}^{(s-1)}, \boldsymbol{n}_{i_m}^{(s)}\right),\label{eq:power}\\
\boldsymbol{x}_{i}^{(s)} &= \text{CONCATENATE}\left(\mathbf{c}_{i}^{(s)},p_{i}^{(s)}\right).
\end{align}
where $\alpha_1^{(s)}(\cdot)$ and $\alpha_2^{(s)}(\cdot)$ are two update functions for channel allocation and power allocation tasks, respectively. 
Following aggregation, two distinct MLPs are employed to process the aggregated messages $\boldsymbol{n}_{i_m}$ and the vertex’s previous feature $\boldsymbol{x}_{i}^{(s-1)}$. The output layer of $\alpha_1^{(s)}(\cdot)$ employs the Softmax function since $\mathbf{c}_{i} = [c_{i}^1,...,c_{i}^{M}]$ is a one-hot vector. We set $c_{i}^{m}$ to 1 if it has the highest value, and all other elements are set to zero. Since there exists a maximum power constraint in the power allocation problem, the Sigmoid function is utilised in the output layer of $\alpha_2^{(s)}(\cdot)$ to normalise the power. The summary of the JCPGNN is presented in Algorithm \ref{alg:cap}.

\begin{algorithm}
\caption{JCPGNN}\label{alg:cap}
\begin{algorithmic}[1]
\State \textbf{Input}: Heterogeneous Graph $\boldsymbol{G}$; node feature matrix $\mathbf{v}_{i_m}$; edge feature matrix $\mathbf{e}_{i_m,j_m}$ and $\mathbf{e}_{i_m,j_n}$; nonlinear transformation $\phi_1(\cdot)$; aggregation function $\phi_2(\cdot)$; update functions $\alpha_1(\cdot)$ and $\alpha_2(\cdot)$;layer number $S$
\State \textbf{Initialisation}: Randomly initialise the weight of functions $\phi_1(\cdot)$, $\phi_2(\cdot)$,$\alpha_1(\cdot)$ and $\alpha_2(\cdot)$; $\boldsymbol{x}_{i}^0 \gets [\mathbf{1},p_{max}]$;
\For{\texttt{$s=1$ to $S$}}
\For{\texttt{$i=1$ to $D$}}
\State Shared message computation step: Gather the messages from neighbours and aggregate them together based on Eq.\eqref{eq:aggreated}.
\State Task-specific step: Calculate the channel $\mathbf{c}_{i}^{(s)}$ and power allocation $p_{i}^{(s)}$ separately based on Eq.\eqref{eq:channel} and \eqref{eq:power}.

\State $\boldsymbol{x}_{i}^{(s)} \gets \text{CONCATENATE}\left(\mathbf{c}_{i}^{(s)},p_{i}^{(s)}\right).$
    \EndFor
\EndFor

\end{algorithmic}
\end{algorithm}

\section{Experiments and Results}\label{sec:experiments}
\subsection{Simulations Setup}\label{sec:simulation}
We adopt the system configuration that encompasses large-scale fading and Rayleigh fading, as detailed in \cite{10283511, liang2019towards}. In this scenario, we consider a system comprising $D$ transceiver pairs situated within a $100 \times 100$ $m^2$ area. The placement of transmitters is randomised within this area, and each receiver is randomly positioned at a distance ranging from 2$m$ to 10$m$ from its corresponding transmitter. The hidden sizes of message computation $\phi_1(\cdot)$ and update functions $\alpha_1(\cdot)$ and $\alpha_2(\cdot)$ are $\{5+M,16,32\}$, $\{33+M,16,8,M\}$ and $\{33+M,16,8,1\}$, respectively. We initialise the power allocation as $\mathbf{P}^{(0)} = [p_{\max},...,p_{\max}]$ and the channel allocation as $\mathbf{C}^{0} =[\mathbf{1},...,\mathbf{1}]^{T}$, where $\mathbf{1} = [1,...,1]\in\mathbb{R}^{1\times M}$ . Since maximising the sum rate means minimising the negative sum rate, the loss function for our graph neural network can be formulated as,
\begin{equation}
\begin{aligned}
L(\theta)= & -\hat{\mathbb{E}}_{\boldsymbol{H}}\Bigg\{\sum_{m=1}^{M}  \sum_{i=1}^{D} R_{i}^m(\mathbf{C,P})\Bigg\}. \\
\end{aligned}
\label{eq:lossfunction}
\end{equation}
 
Where $\hat{\mathbb{E}}$ denotes the expectation with respect to the empirical distribution of the channel samples and 
$\boldsymbol{H}$ is the channel matrix including all the CSI between pairs.
Here, we train our networks under 10000 training samples and test the performance with 1000 testing samples.

To assess the effectiveness of our proposed JCPGNN, we consider four baselines for performance comparison. The optimal solution is derived through an exhaustive search for channel allocation, with the state-of-the-art algorithm for power allocation (WMMSE\cite{shi2011iteratively}) employed for power allocation in each iteration. As a baseline, we utilise the GNN structure proposed in \cite{chen2021gnn} for channel allocation. However, due to the modification in our setup, allowing one channel to be shared by multiple D2D pairs (as opposed to their constraint of one channel shared by at most one D2D pair), we adapted their algorithm and incorporated WMMSE for power allocation. The four baselines are listed below,

\begin{itemize}
    \item Optimal: The optimal solution obtained through the exhaustive search for channel allocation and WMMSE for power allocation.
    \item JCPGNN: Our proposed joint channel and power allocation GNN algorithm, featuring shared message computation and two task-specific output layers.
    \item RR: A traditional optimisation algorithm involving Round-robin for channel allocation \cite{luong2008channel} and GNN for power allocation.
    \item GCA: GNN-based channel allocation algorithm \cite{chen2021gnn}, then followed by WMMSE for power allocation.
    \item Closest: The algorithm allocates any two closest pairs to different channels and subsequently utilises WMMSE for power allocation. 
\end{itemize}
Given that the Closest algorithm generates an unbalanced channel allocation, resulting in a varying number of transceiver pairs in each training sample, it introduces a challenge when employing GNN architectures for power allocation. Many GNN models assume a fixed number of input nodes during training, making it less suitable for scenarios with varying numbers of transceiver pairs. Therefore, for power allocation, we refrain from considering the use of GNN in such cases.
\subsection{Performance Comparison}\label{subsec:performance}
\begin{figure}[htbp]
\centerline{\includegraphics[width=9cm]{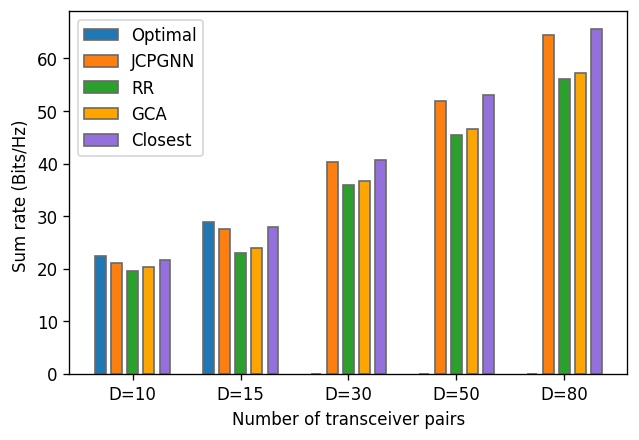}}
\caption{The average sum rate for two channels scenario.}
\label{fig:performance_2C}
\end{figure}
To validate the effectiveness of our proposed algorithm, we conduct experiments under a two-channel scenario. The results for sum-rate performance under different numbers of transceiver pairs are illustrated in Figure \ref{fig:performance_2C}. Notably, due to the high computational complexity, the optimal solution's performance is presented only up to $D=15$. Our observations reveal that the proposed JCPGNN achieves $95\%$ of the optimal solution's performance for $D=10$ and $D=15$. Additionally, our framework outperforms all other baseline algorithms when the number of transceiver pairs is small and achieves comparable performance to the Closest algorithm as the number of transceiver pairs increases.

The superior performance of JCPGNN compared to RR and GCA algorithms can be attributed to the sharing and exchanging of information between power allocation and channel allocation tasks within JCPGNN. In contrast, GCA employs WMMSE power allocation in the postprocessing steps of channel allocation. While JCPGNN slightly underperforms the Closest algorithm in cases with a large number of transceiver pairs, this can be attributed to the elevated interference levels. The Closest algorithm effectively mitigates interference, leading to improved performance. 

\subsection{Generalisation Capability}\label{sec:gen}
\begin{table}[]
\centering
\caption{Generalisation performance of JCPGNN to larger area}
\begin{tabular}{|c|c|c|c|c|c|}
\hline \multicolumn{2}{|c|}{ System Scales } & \multicolumn{1}{c|}{$D=15$} & $D=30$ & $D=50$ & $D=80$ \\
\hline \multirow{3}{*}{$M=2$} & JCPGNN & 29.38 & 52.46 & 82.48 & 123.86 \\
\cline { 2 - 6 } & RR & 25.17 & 47.45 & 75.01 & 113.64\\
\cline { 2 - 6 }  & Closest & 29.54 & 54.10 & 86.25 & 131.78\\
\hline \multirow{3}{*}{$M=3$} & JCPGNN & 33.90 & 61.25 & 96.20 & 141.40\\
\cline { 2 - 6 } & RR & 31.59& 56.41 & 90.07 & 131.18 \\
\cline { 2 - 6 } & Closest & 33.16 & 62.23 & 97.70 & 150.37 \\
\hline 
\end{tabular}
\label{Tab:generalise_area}
\end{table}

Apart from achieving high sum rate performance, being able to generalise to larger scale problems is also important. In this section, we investigate the generalisation capability of JCPGNN.
We first train JCPGNN and RR on a small network, e.g., $D=10$ within a $100 \times 100$ $m^2$ area. Then we fix the density of transceiver pairs and increase the size of the region. The generalisation performance is shown in Table~~\ref{Tab:generalise_area}. We observe that the performance of JCPGNN attains an impressive $94\%$ of the state-of-the-art traditional algorithm's performance, even as the network size increases by a factor of 8. This underscores the robust generalisation capabilities of JCPGNN across varying network scales.

\subsection{Robustness to Corrupted CSI}
\begin{figure}[htbp]
\centerline{\includegraphics[width=9cm]{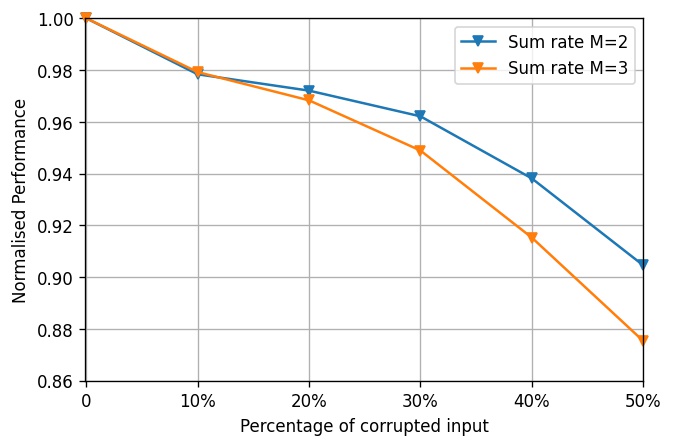}}
\caption{Normalised performance of JCPGNN with difference percentage of corrupted CSI input. }
\label{fig:corrupted}
\end{figure}
Since it is hard to estimate the real CSI in dynamic wireless networks, to test the robustness of the proposed JCPGNN algorithm, we apply it with corrupted input CSI. We train the proposed JCPGNN with full CSI and test it with different percentages of corrupted CSI. The performance of the JCPGNN for two-channel and three-channel scenarios when $D=80$ is shown in Figure~\ref{fig:corrupted}. Note that the performance is normalised by the performance of JCPGNN achieved with full CSI. Here, $10\%$ corrupted CSI input means removing $10\%$ of total CSI when training the graph neural networks. We can observe from the figure that the proposed JCPGNN can still achieve $90.5\%$ and $87.5\%$ of optimal performance for two and three-channel scenarios, respectively when $50\%$ of the CSI is missing. This robustness feature is desirable in practical wireless IoT networks where some CSI may be unavailable.

\subsection{Time Complexity}
\begin{figure}[htbp]
\centerline{\includegraphics[width=9cm]{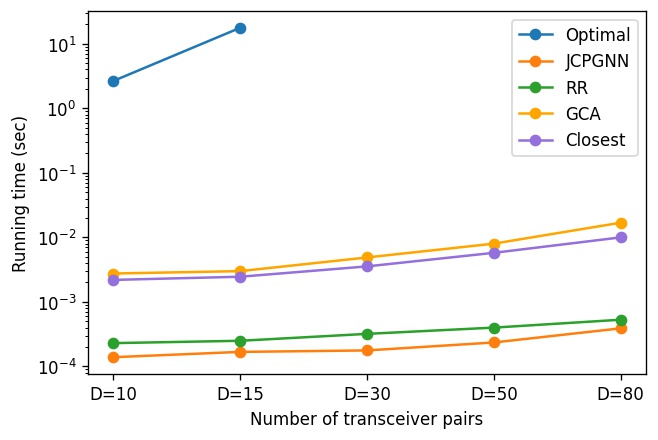}}
\caption{Average running time for all the algorithms with a two-channel scenario (the figure is presented on a logarithmic scale).}
\label{fig:timecomplexity}
\end{figure}

The average running time for the algorithms under the same experimental setting as depicted in Figure~\ref{fig:performance_2C} is illustrated in Figure~\ref{fig:timecomplexity}. Due to the high complexity, we are not able to generate optimal solutions for larger $D$. Our observation indicates that JCPGNN exhibits significantly lower complexity compared to other benchmarks, primarily attributable to the inefficiency of the WMMSE algorithm. The marginally increased computational time observed for the RR algorithm, compared to the proposed JCPGNN, can be attributed to the fact that test samples pass through the GNN twice in the RR algorithm for obtaining power allocation of two channels separately. The running time of the exhaustive search method grows exponentially with the problem size,  while the increase for JCPGNN is infinitesimal. For instance, by selecting $D =15$, JCPGNN achieves a $95\%$ optimal performance while reducing the required running time by roughly $10^5$ times compared to the exhaustive search method. Compared to traditional Closest algorithms, JCPGNN can achieve similar performance with only $4\%$ of its running time. The complexity of JCPGNN can be further reduced by pruning the edge of graph representation\cite{chen2023accelerating}.  

\section{conclusion}

In this study, we address the challenges of resource allocation in wireless communication networks, focusing on the joint optimisation of channels and power resources. Leveraging a graph-based representation, we propose an innovative JCPGNN model, that incorporates a shared message computation layer and two task-specific layers, dedicated to channel and power allocation tasks, respectively. Through extensive experiments, we validate the performance of our algorithm, comparing it with traditional optimisation methods. Notably, our results showcase that the JCPGNN achieves comparable performance while exhibiting superior computational efficiency, highlighting its potential for real-time implementation in wireless communication networks. Additionally, the demonstrated generalisability of the GNN framework to various network configurations and its robustness to corrupted input features further emphasise the flexibility and reliability of our proposed solution.
\bibliographystyle{IEEEtran}
\bibliography{main}

\begin{thebibliography}{10}
\providecommand{\url}[1]{#1}
\csname url@samestyle\endcsname
\providecommand{\newblock}{\relax}
\providecommand{\bibinfo}[2]{#2}
\providecommand{\BIBentrySTDinterwordspacing}{\spaceskip=0pt\relax}
\providecommand{\BIBentryALTinterwordstretchfactor}{4}
\providecommand{\BIBentryALTinterwordspacing}{\spaceskip=\fontdimen2\font plus
\BIBentryALTinterwordstretchfactor\fontdimen3\font minus \fontdimen4\font\relax}
\providecommand{\BIBforeignlanguage}[2]{{%
\expandafter\ifx\csname l@#1\endcsname\relax
\typeout{** WARNING: IEEEtran.bst: No hyphenation pattern has been}%
\typeout{** loaded for the language `#1'. Using the pattern for}%
\typeout{** the default language instead.}%
\else
\language=\csname l@#1\endcsname
\fi
#2}}
\providecommand{\BIBdecl}{\relax}
\BIBdecl

\bibitem{feng2013device}
D.~Feng, L.~Lu, Y.~Yuan-Wu, G.~Y. Li, G.~Feng, and S.~Li, ``Device-to-device communications underlaying cellular networks,'' \emph{IEEE Transactions on communications}, vol.~61, no.~8, pp. 3541--3551, 2013.

\bibitem{meshgi2015joint}
H.~Meshgi, D.~Zhao, and R.~Zheng, ``Joint channel and power allocation in underlay multicast device-to-device communications,'' in \emph{2015 IEEE International Conference on Communications (ICC)}.\hskip 1em plus 0.5em minus 0.4em\relax IEEE, 2015, pp. 2937--2942.

\bibitem{zhao2015resource}
W.~Zhao and S.~Wang, ``Resource sharing scheme for device-to-device communication underlaying cellular networks,'' \emph{IEEE transactions on communications}, vol.~63, no.~12, pp. 4838--4848, 2015.

\bibitem{abdallah2018power}
A.~Abdallah, M.~M. Mansour, and A.~Chehab, ``Power control and channel allocation for {D2D} underlaid cellular networks,'' \emph{IEEE Transactions on Communications}, vol.~66, no.~7, pp. 3217--3234, 2018.

\bibitem{gao2022decentralized}
Z.~Gao, Y.~Shao, D.~Gunduz, and A.~Prorok, ``Decentralized channel management in {WLAN}s with graph neural networks,'' \emph{arXiv preprint arXiv:2210.16949}, 2022.

\bibitem{10283511}
L.~Chen, J.~Zhu, and J.~Evans, ``Graph neural networks for power allocation in wireless networks with full duplex nodes,'' in \emph{2023 IEEE International Conference on Communications Workshops (ICC Workshops)}, 2023, pp. 277--282.

\bibitem{he2020resource}
Z.~He, L.~Wang, H.~Ye, G.~Y. Li, and B.-H.~F. Juang, ``Resource allocation based on graph neural networks in vehicular communications,'' in \emph{GLOBECOM 2020-2020 IEEE Global Communications Conference}.\hskip 1em plus 0.5em minus 0.4em\relax IEEE, 2020, pp. 1--5.

\bibitem{nakashima2020deep}
K.~Nakashima, S.~Kamiya, K.~Ohtsu, K.~Yamamoto, T.~Nishio, and M.~Morikura, ``Deep reinforcement learning-based channel allocation for wireless {LAN}s with graph convolutional networks,'' \emph{IEEE Access}, vol.~8, pp. 31\,823--31\,834, 2020.

\bibitem{chen2021gnn}
T.~Chen, X.~Zhang, M.~You, G.~Zheng, and S.~Lambotharan, ``A {GNN}-based supervised learning framework for resource allocation in wireless iot networks,'' \emph{IEEE Internet of Things Journal}, vol.~9, no.~3, pp. 1712--1724, 2021.

\bibitem{shen2020graph}
Y.~Shen, Y.~Shi, J.~Zhang, and K.~B. Letaief, ``Graph neural networks for scalable radio resource management: Architecture design and theoretical analysis,'' \emph{IEEE Journal on Selected Areas in Communications}, vol.~39, no.~1, pp. 101--115, 2020.

\bibitem{hornik1989multilayer}
K.~Hornik, M.~Stinchcombe, and H.~White, ``Multilayer feedforward networks are universal approximators,'' \emph{Neural networks}, vol.~2, no.~5, pp. 359--366, 1989.

\bibitem{zhang2022learning}
X.~Zhang, Z.~Zhang, and L.~Yang, ``Learning-based resource allocation in heterogeneous ultradense network,'' \emph{IEEE Internet of Things Journal}, vol.~9, no.~20, pp. 20\,229--20\,242, 2022.

\bibitem{liang2019towards}
F.~Liang, C.~Shen, W.~Yu, and F.~Wu, ``Towards optimal power control via ensembling deep neural networks,'' \emph{IEEE Transactions on Communications}, vol.~68, no.~3, pp. 1760--1776, 2019.

\bibitem{shi2011iteratively}
Q.~Shi, M.~Razaviyayn, Z.-Q. Luo, and C.~He, ``An iteratively weighted {MMSE} approach to distributed sum-utility maximization for a {MIMO} interfering broadcast channel,'' \emph{IEEE Transactions on Signal Processing}, vol.~59, no.~9, pp. 4331--4340, 2011.

\bibitem{luong2008channel}
T.-T. Luong, B.-S. Lee, and C.~K. Yeo, ``Channel allocation for multiple channels multiple interfaces communication in wireless ad hoc networks,'' in \emph{NETWORKING 2008 Ad Hoc and Sensor Networks, Wireless Networks, Next Generation Internet: 7th International IFIP-TC6 Networking Conference Singapore, May 5-9, 2008 Proceedings 7}.\hskip 1em plus 0.5em minus 0.4em\relax Springer, 2008, pp. 87--98.

\bibitem{chen2023accelerating}
L.~Chen, J.~Zhu, and J.~Evans, ``Accelerating graph neural networks via edge pruning for power allocation in wireless networks,'' \emph{arXiv preprint arXiv:2305.12639}, 2023.

\end{thebibliography}

\vspace{12pt}

\end{document}